\documentclass[twocolumn,aps,prl,showpacs,floatfix,superscriptaddress]{revtex4}

\usepackage{amsmath}    
\usepackage{graphicx}   
\usepackage{epsfig}
\usepackage{verbatim}   
\usepackage{color}      
\usepackage{subfigure}  
\usepackage{hyperref}   

\begin{document}

\def\bx{{\bf x}}
\def\bu{{\bf u}}
\def\bv{{\bf v}}
\def\bk{{\bf k}}
\def\bp{{\bf p}}
\def\bq{{\bf q}}
\def\br{{\bf r}}
\def\eps{\varepsilon}
\def\epsCr{\varepsilon_\mathrm{cr}}

\title{Turbulence on a Fractal Fourier Set\footnote{postprint version published open access on 
Phys. Rev. Lett. {\bf 115} 264502 (2015)}} \author{Alessandra
  S. Lanotte} \affiliation{ISAC-CNR and INFN Sez. Lecce, 73100 Lecce,
  Italy} \email{Contact author: a.lanotte@isac.cnr.it}

\author{Roberto Benzi}
\affiliation{Dept. of Physics and INFN, University of Rome Tor Vergata, Via della Ricerca Scientifica 1, 00133 Roma, Italy}

\author{Shiva K. Malapaka}
\affiliation{Dept. of Physics and INFN, University of Rome Tor Vergata, Via della Ricerca Scientifica 1, 00133 Roma, Italy, and IIIT- Bangalore, Electronics City, Hosur Road, Bangalore 560100, India}

\author{Federico Toschi} \affiliation{Department of Applied Physics,
  Eindhoven University of Technology, 5600 MB Eindhoven, 
  Netherlands, and IAC CNR, Via dei Taurini 19, 00185 Roma, Italy}

\author{Luca Biferale}
\affiliation{Dept. of Physics and INFN, University of Rome Tor Vergata, Via della Ricerca Scientifica 1, 00133 Roma, Italy}

\begin{abstract}
A novel investigation of the nature of intermittency in
incompressible, homogeneous and isotropic turbulence is performed by a
numerical study of the Navier-Stokes equations constrained on a
fractal Fourier set. The robustness of the energy transfer and of the
vortex stretching mechanisms is tested by changing the fractal
dimension, $D$, from the original three dimensional case to a strongly
decimated system with $D=2.5$, where only about $3\%$ of the Fourier
modes interact. This is a unique methodology to probe the statistical
properties of the turbulent energy cascade, without breaking any of
the original symmetries of the equations. While the direct energy
cascade persists, deviations from the Kolmogorov scaling are observed
in the kinetic energy spectra. A model in terms of a correction with a
linear dependency on the codimension of the fractal set, $E(k) \sim
k^{-5/3 + 3-D}$, explains the results. At small scales, the
intermittency of the vorticity field is observed to be quasisingular
as a function of the fractal mode reduction, leading to an almost
Gaussian statistics already at $D \sim 2.98$. These effects must be
connected to a genuine modification in the triad-to-triad nonlinear
energy transfer mechanism.
\end{abstract}

\date{} \pacs{ } \maketitle
\begin{figure}
\includegraphics[width=8.2cm,scale=0.52]{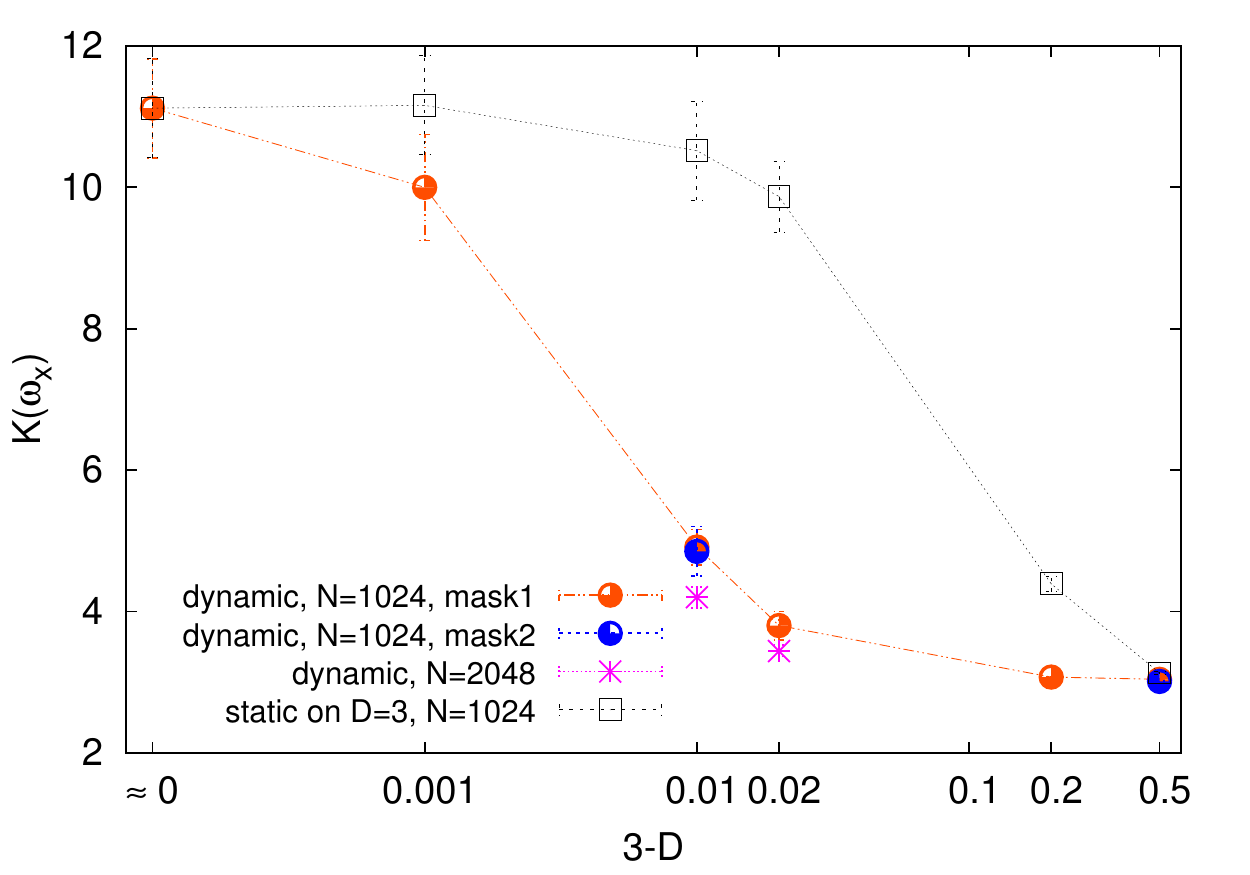}
\vspace{-0.5cm}
\caption{Lin-log plot of vorticity kurtosis vs the dimension deficit
  $3-D$. The upper curve is obtained from the application of the {\it
    a posteriori} static mask on $D=3$ velocity field snapshots. The
  lower curve comes from the Fourier decimated DNS. Data for $D=2.99$
  and $D=2.5$, with a different realization of the fractal
  projection, are indistinguishable. Data at resolution $N=2048$ are
  also plotted.}
\label{fig:5}
\end{figure}

\begin{figure}
\includegraphics[width=8.2cm,scale=0.52]{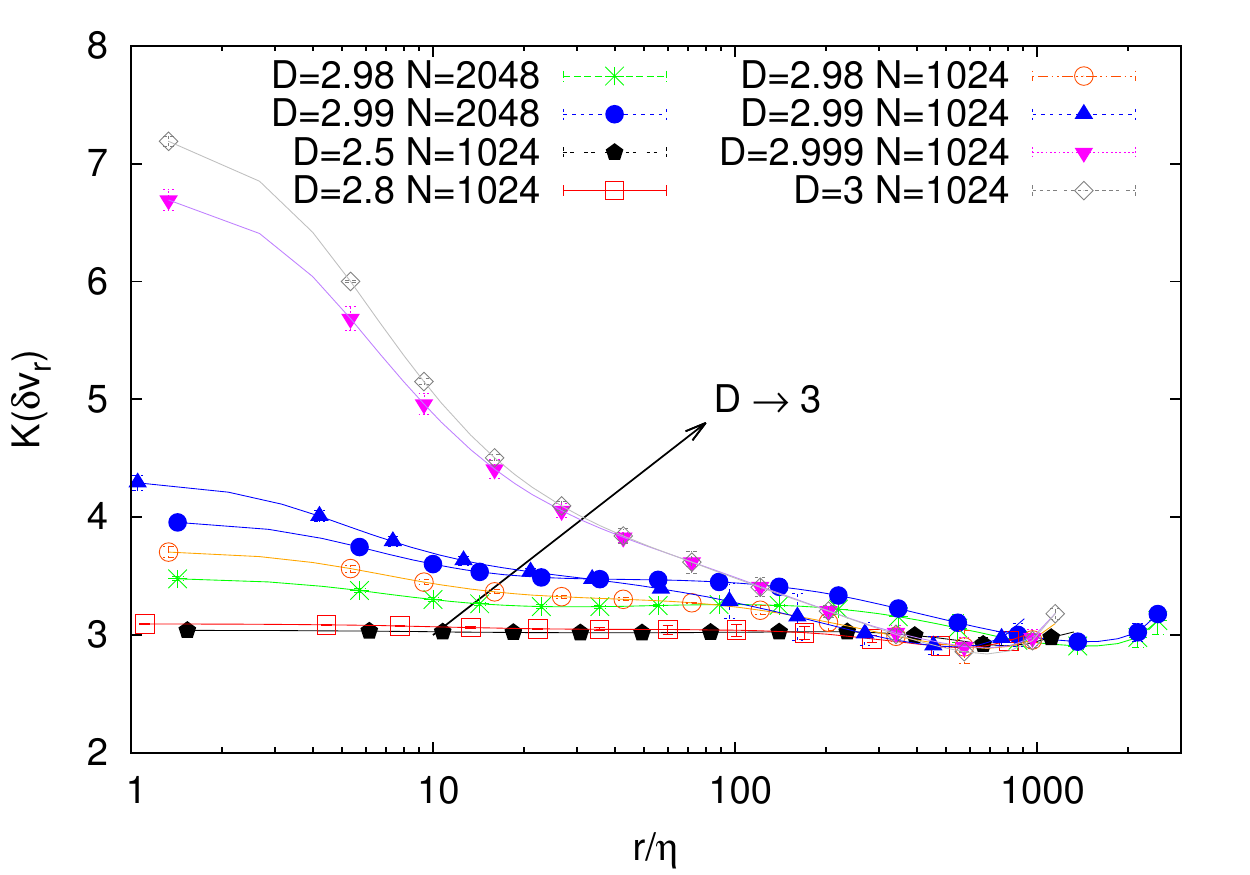}
\vspace{-0.5cm}
\caption{Lin-log plot of longitudinal velocity increment kurtosis
  $K(\delta v_r)$ vs the scale separation normalised by the Kolmogorov
  scale, $r/\eta$. For all data sets, error bars are within the symbol
  width.}
\label{fig:6}
\end{figure}

 Understanding and controlling the energy
transfer process in turbulent flows is a key problem for a broad range
of fields, such as astrophysics \cite{biskamp}, atmospheric or ocean
sciences \cite{pedlosky}, mathematics and engineering
\cite{pope_book}. The main obstacle hampering theoretical, numerical,
and phenomenological advancements is {\it intermittency}: the presence
of strong non-Guassian and out-of-equilibrium velocity fluctuations in
a wide range of scales \cite{frisch,Kr71,Kr74,MK00,SreFa06}. The
energy cascade is fully chaotic, nonlinear and driven by the
vortex-stretching mechanism, i.e., the tendency of the flow to amplify
vorticity in thin, long filaments. A long debate exists whether or not
the presence of such geometrical structures is correlated to the
non-Gaussian statistics observed at small scales \cite{frisch}. Many
authors have focused on a vortex-by-vortex analysis, looking for the
signatures of quasisingularities or extreme events due to specific
dynamical properties of the Navier-Stokes (NS) equations
\cite{chorin,pullin,PPSAM95,Tsi97,Pi99,Luthi05,Ka15}. Other approaches
are based on a Fourier description, such as closure
\cite{K61,orszag77} and renormalization-group theories
  \cite{DDM,eyink,antonov}.\\ In this Letter, we investigate the
origin of intermittency in turbulent flows by a novel strategy. The
idea consists of modifying the nonlinear interactions of
Navier-Stokes equations, without introducing extra forces and without
breaking the symmetries of the original equations, such as statistical
homogeneity, isotropy and rescaling properties in the inviscid
limit. To achieve this, we adopt a recently proposed numerical
methodology \cite{frisch2012} to solve NS equations on a preselected,
multiscale (fractal) set of Fourier modes. This allows us to control
the number of degrees of freedom participating in the non-linear dynamics
by changing one free parameter only, the fractal dimension of the
Fourier set, $D$. For $D=3$, the original problem is recovered. In the
sequel, we describe the methodology, the numerical setup and the main
results concerning the quasisingular effect of fractal mode reduction
on the small-scale intermittency.  \\
\noindent {\it Methodology}. Fractal mode reduction is realized via
the Fourier decimation operator ${\cal P}^{D}$, acting in the space of
divergece-free velocity fields as follows \cite{frisch2012}. We
  define $\bv(\bx,t)$ and $\bu(\bk,t)$ as the real and Fourier space
  representation of the velocity field in $D=3$, respectively. The
  decimated field, $\bv^{D}(\bx,t)$, is obtained as:
\begin{equation}
\label{eq:decimOper}
{\bf v}^{D}(\bx,t)= {\cal P}^{D}{\bf v}(\bx,t)=\hspace{-1mm}
\sum_{{\bf k}\in {\cal Z}^3}\hspace{-1mm} e^{i {\bf k \cdot x}}\,\gamma_{\bf k}\bu(\bk,t)\,.
\end{equation}
The random numbers $\gamma_{\bk}$ are quenched in time and are:
\begin{equation}
\label{eq:theta}
\gamma_{\bf k} =
\begin{cases}
1, & \text{with probability}\ h_k\,, \\
0, & \text{with probability}\ 1-h_k, k\equiv|{\bf k}|\,.
\end{cases}
\end{equation}
The choice for the probability $h_k \propto (k/k_0)^{D-3}$, with $0< D
\le 3$ ensures that the dynamics is isotropically decimated to a
$D$-dimensional Fourier space. The factors $h_k$ are chosen
independently and preserve Hermitian symmetry $\gamma_k= \gamma_{-k}$
so that ${\cal P}^{D}$ is self-adjoint. The NS equations for the
velocity field decimated on a fractal Fourier set are then defined
as:
\begin{equation}
\label{eq:decimNS}
\partial_t {\bf v}^{D} = {\cal P}^{D}N({\bf v}^{D},{\bf v}^{D}) + 
  \nu \,\nabla^2 {\bf v}^{D} + {\bf f}^{D}\,. 
\end{equation}
At each iteration the nonlinear term, $N({\bf v},{\bf v}) = - {\bf
  v}\cdot {\bf {\nabla v}} - {\bf \nabla}p$, is projected on the
quenched fractal set, to constrain the dynamical evolution to evolve
on the same Fourier skeleton at all times. Similarly, the initial
condition and the external forcing must have a support on the same set
of Fourier modes. Let us notice that the projection acts as a
self-similar Galerkin truncation. In the $(L^2)$ norm, $\|{\bf v}\|
\propto \int |{\bf v}({\bf x})|^2 d^3 x$, the self-adjoint operator
${\cal P}^{D}$ commutes with the gradient and viscous operators. Since
${\cal P}^{D} {\bf v}^{D}= {\bf v}^{D}$, it then follows that both
energy and helicity are conserved in the inviscid and unforced limit,
exactly as in the original problem with $D=3$. As a result of the
Fourier decimation, the velocity field is embedded in a three
dimensional space, but effectively possesses a number of Fourier modes
that grows slower with decreasing $D$. The degrees of freedom inside a
sphere of radius $k$ go as $\#_{dof}(k) \sim k^{D}$.\\ This idea,
introduced for the first time in \citep{frisch2012}, has been used to
test the hypothesis that two-dimensional turbulence in the inverse
energy cascade approaches a quasiequilibrium state
\citep{lvov}. Fourier decimation methods are not new for
hydrodynamics: we mention protocols with a specific degree of mode
reduction \citep{GLR96,MPPZ96,DLE07}, and the extreme truncation
criterion of shell models for the turbulent energy cascade
\cite{bif03}. Results are puzzling. For NS equations at small Reynolds
numbers \cite{GLR96}, intermittency strongly depends on the amount of
scales resolved in the inertial range. However, in the case of
shell models, intermittency is observed to be a function of the {\it
  effective dimension} \cite{giuliani}, and coinciding with the one
measured in the original NS equations \cite{paladin-vulpiani}, when
energy and helicity are the two inviscid invariants. Note that fractal
Fourier mode-reduction changes also the relative population of
local-to-nonlocal triadic interactions \cite{Kr71}, since triads with
all modes in the high wave number range have a larger probability to be
decimated. Furthermore, being an exquisitely dynamical approach, it
  is different from {\it a posteriori} filtering techniques, largely
  exploited to analyze turbulent data \cite{farge}. \\ A
  pseudospectral method is adopted to solve Eqs.~(\ref{eq:decimNS})
  in a periodic box of length $L=2\pi$ at resolution $N=1024^3$ and
  $2048^3$, dealiased with the two-thirds rule; time stepping is
  implemented with a second-order Adams-Bashforth scheme. A
  large-scale forcing \cite{pope} keeps the total kinetic energy
  constant in a range of shells, $ 0.7 \le |{\bf k}| < 1.7$, leading
  to a steady, homogeneous and isotropic flow. We performed direct
  numerical simulation (DNS) runs by changing the fractal dimension
  $2.5 \le D \le 3$, the spatial resolution, the viscosity and the
  realization of the fractal, quenched mask. Table \ref{table:param}
  summarizes the relevant parameters.\\
\begin{table}
\begin{center}
\begin{tabular}{lllllllll}
\hline
$D$ & $3$ & $2.999$ & $2.99$ & $2.99$ & $2.98$ & $2.98$ & $2.8$ & $2.5$\\ 
$N$  & $1024$ & $1024$  & $1024$ & $2048$ & $1024$ & $2048$ & $1024$& $1024$ \\
$M_r$ & $100\%$ & $99\%$  & $93\%$ & $92\%$ & $87\%$ & $85\%$ & $25\%$& $3\%$ \\ 
$\eta$ & $0.75$ & $0.75$  & $0.95$ & $0.70$ & $0.75$ & $0.70$ & $0.90$& $0.65$ \\ 
${\cal N}_T$ & 10 & 10 & 11 & 10 & 11 & 7 & 10 & 20 \\ 
$E$    & $3.1$   & $3.1$   & $3.1$   & $3.3$   & $3.3$   & $3.5$      & $4.1$  & $5.4$ \\
$\nu \cdot 10^{4}$ & $6.0$ & $6.0$ & $6.0$ & $2.0$ & $6.0$ & $2.0$ & $6.0$ & $1.5$\\
$Re \cdot 10^{-3}$  & $3.9$  & $3.9$ & $3.8$ & $11.8$ & $3.9$ & $12.1$ & $4.0$ & $15.4$ \\
\hline
\end{tabular}
\caption{DNS parameters. $D$: fractal dimension; $N$: grid resolution;
  $M_r$: percentage of surviving Fourier modes; $\eta$: Kolmogorov
  length scale in unit of the grid spacing $\Delta x= L/N$; ${\cal
    N}_{T}$: number of large-scale eddy-turnover-times in the steady
  state; $E$: total kinetic energy $\langle v^2 \rangle /2$; $\nu$:
  viscosity; $Re$: Reynolds number $Re=E^{1/2} {\cal L}/\nu$, where
  the integral scale ${\cal L}$ is estimated from the kinetic energy
  spectrum.}
\label{table:param}
\end{center}
\end{table}
\noindent{\it Results}. The starting point of our analysis is the
shell-to-shell energy transfer in the Fourier space. Following the
notation adopted in Ref.~\cite{Kr71}, we write the energy spectrum for
a generic flow in dimension $D$ as:
\begin{equation}
E^{D}(k) = \int_{|\bk_1|=k} d^3k_1 \gamma_{\bk_1} \int d^{3}k_2 \gamma_{\bk_2} \langle \bu(\bk_1)\, \bu(\bk_2) \rangle\,,
\label{eq:en_balance}
\end{equation}
where the decimation factor $\gamma_{\bk}$ takes into account that the
Fourier mode ${\bf k}$ is active with probability $h_{\bf
  k}$. Similarly, we can write for the energy flux across a Fourier
mode $k$, $\Pi^{D}(k)= \int_{|\bk_1|< k} d^3k_1 \partial_t E(k_1)$:
\begin{equation}
\Pi^{D}(k)=\int_{|\bk_1|<k}d^3k_1\gamma_{\bk_1}\int d^3k_2 d^3k_3 \gamma_{\bk_2}\gamma_{\bk_3} S(\bk_1|\bk_2,\bk_3),
\label{eq:en_transfer}
\end{equation}
where the explicit form of the symmetric triadic correlation function
is \cite{RoseSulem}: $S(\bk_1|\bk_2,\bk_3) = -Im [\langle (\bk_1\cdot
  \bu(\bk_3))( \bu(\bk_1)\cdot \bu(\bk_2))\rangle + \langle
  (\bk_1\cdot \bu(\bk_2))( \bu(\bk_1)\cdot
  \bu(\bk_3))\rangle]$. Supposing a power-law behavior of the
velocity fluctuations $u(k) \sim k^{-a}$, we can predict a
self-similar scaling of the energy flux as $\Pi^{D}(\lambda k) \sim
\lambda^{3D+1-3a} \Pi^{D}(k)$. In this expression, the rescaling
factor $\lambda^{3D}$ is due to the integral over the variables
$(\bk_1,\bk_2,\bk_3)$, while $\lambda^{1-3a}$ comes from the triadic
nonlinear term. \\If a constant energy flux develops in the inertial
range of scales, the following dimensional relation holds:
\begin{equation}
\label{eq:spectrum}
a = D+1/3 \rightarrow E^{D}(k) \sim k^{3-D}\,E^{K41}(k) \,,
\end{equation}
where $E^{K41}(k) \sim k^{-5/3}$ is the Kolmogorov 1941 spectrum
expected for the original case with $D=3$. In the previous dimensional
argument, the tiny intermittent corrections to the spectrum scaling
exponent are neglected \cite{IsGoKa09}, while prefactors are omitted
for simplicity. The relation~(\ref{eq:spectrum}) is obtained by
noticing that, because of homogeneity, we have that $\langle\,
\bu(\bk_1) \cdot \bu(\bk_2)\, \rangle \propto F(\bk_1)
\delta^{3}(\bk_1+\bk_2)$, and by also noticing that the decimation
projector verifies the identity $ (\gamma_{\bk})^2 = \gamma_{\bk}$.
\begin{figure}
\vspace{-0.7cm}
\includegraphics[width=8.2cm,scale=0.52]{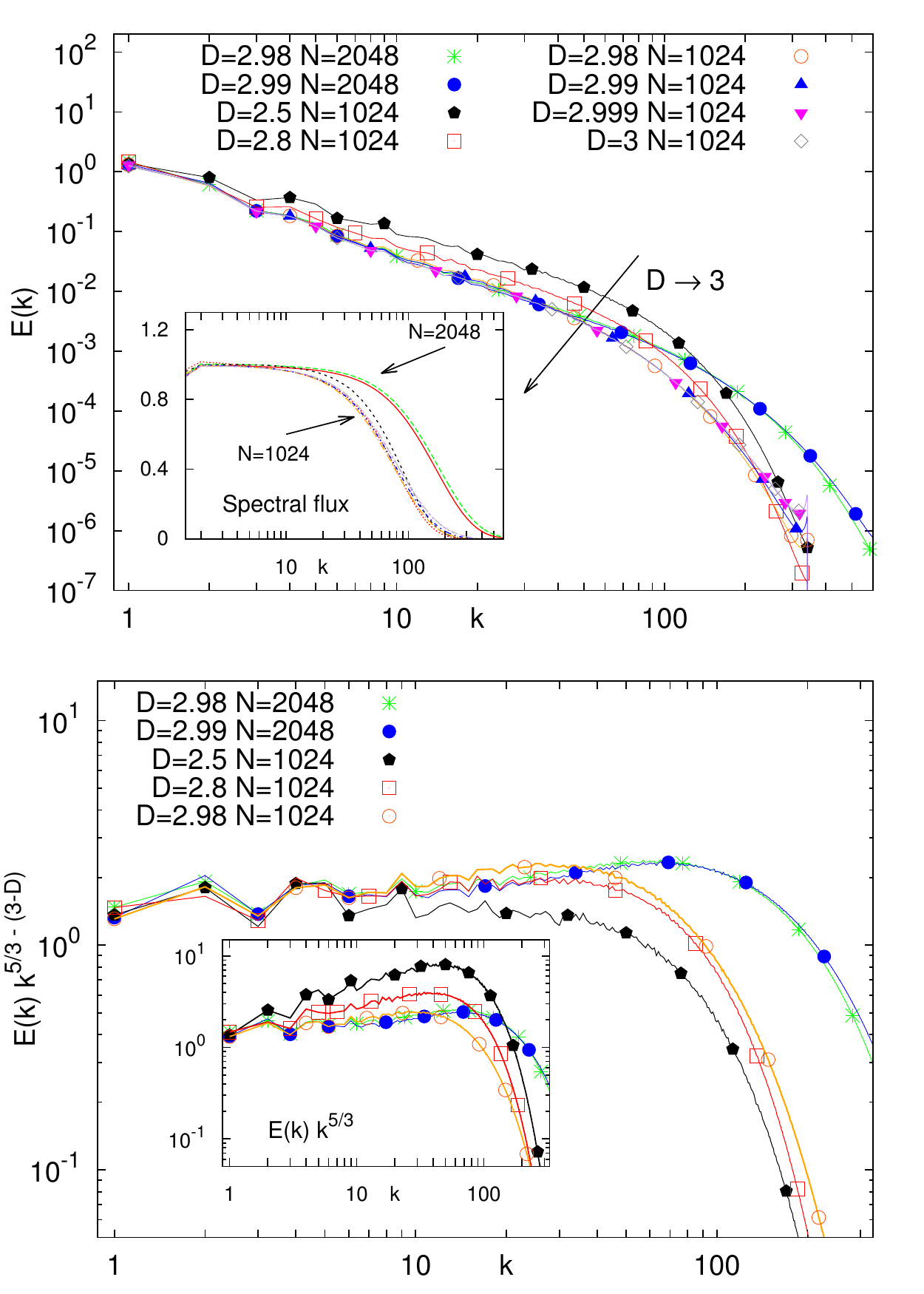}
\caption{(Top): Log-Log plot of the mean kinetic energy spectra for
  different $D$; in the inset, the mean kinetic energy
  fluxes. (Bottom): Compensated energy spectra $E^{D}(k)\, k^{5/3-3+D}$
  vs the wave number $k$; in the inset, the compensation is done with
  the Kolmogorov prediction, $E^{D}(k) \,k^{5/3}$.}
\label{fig:2}
\end{figure}
As a result, the dynamical effect of Fourier decimation is to make the
energy spectrum shallower than the Kolmogorov one for
three-dimensional turbulence, also predicting the existence of a
critical dimension $D=7/3$, when the spectrum becomes ultraviolet
divergent \cite{frisch} in the limit of zero viscosity. By decreasing
$D$ in the presence of a forward energy cascade, the system has fewer
modes available to transfer the same amount of energy~(see Table
\ref{table:param}), and the velocity field becomes increasingly
rougher.\\ In the upper panel of Fig.~\ref{fig:2}, we plot the
energy spectra and the associated energy fluxes, for all DNS. It shows
that when increasing the grid resolution for fixed $D$, from $N=1024$
to $N=2048$, no appreciable differences are observed, indicating that
the presence of a forward energy cascade appears robust and Reynolds
independent. In the lower panel we also show that the spectra
compensate well with the prediction (\ref{eq:spectrum}), while they
fail to satisfactorily compensate with the classical K41 prediction
when $D<3$. We stress that the latter result is significant for
$D=2.5$ and $D=2.8$ only. For the other dimensions $D$, the effect is
so small that it might fall within the intermittent correction of the
original Navier-Stokes case at $D=3$.  Moreover, the effect of the
quenched disorder is robust: spectra obtained with different
realizations of the mask do not show any statistically significant
difference (not shown). \\ Figure~\ref{fig:2} (upper inset) shows
that, when decreasing the fractal dimension $D$, the mean energy
transfer towards small scales is almost unchanged, i.e. the hypothesis
leading to the relation (\ref{eq:spectrum}) is well verified. On the
other hand, temporal fluctuations of the energy flux increase with $D$
(not shown).\\ It might be argued that the effect of fractal Fourier
decimation is purely geometrical and that the main dynamical processes
are unchanged. To show that this is not the case, it is useful to
analyze the effect of a {\it static} Fourier decimation.
\begin{figure}[t]
\includegraphics[width=8.2cm,scale=0.52]{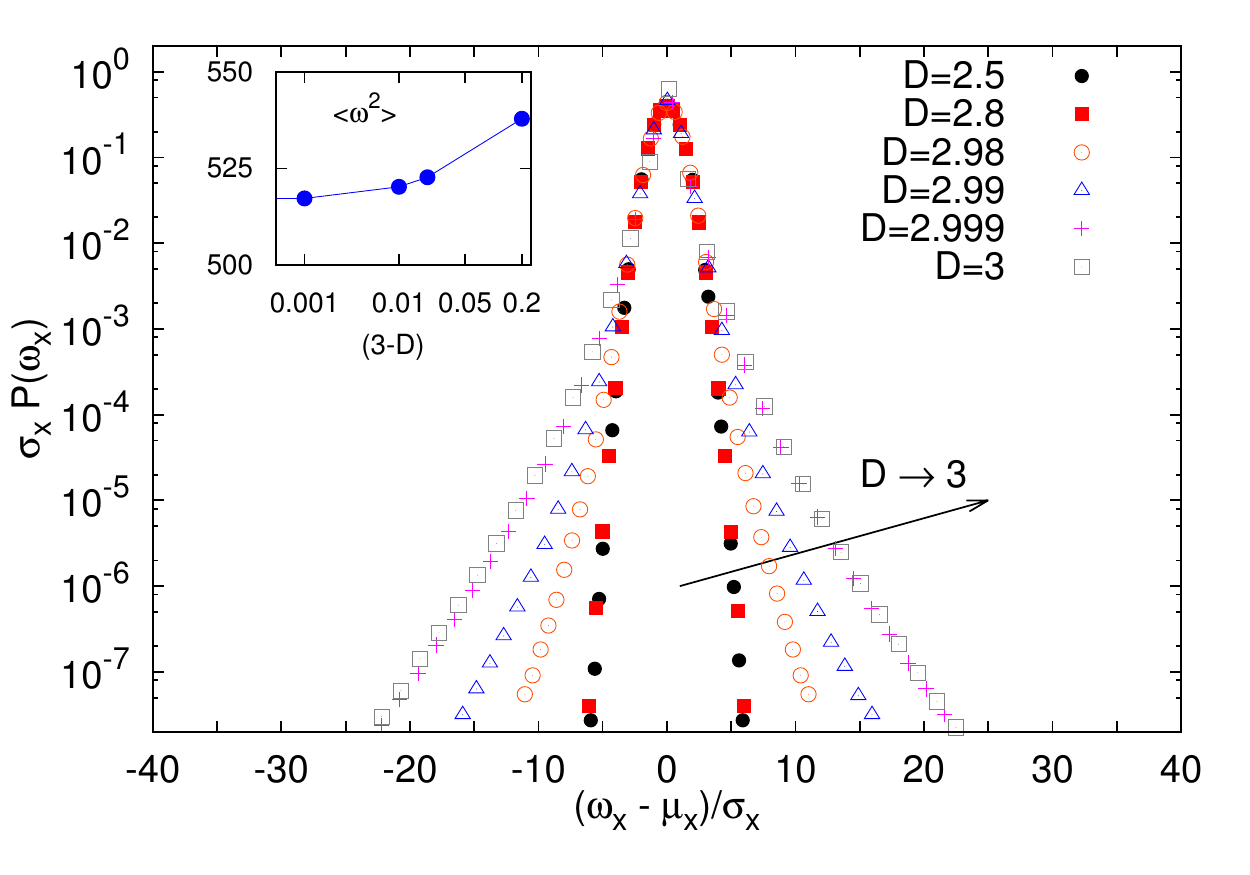}
\vspace{-0.5cm}
\caption{Probability density function of the vorticity component
  $\omega_x$, normalized to its standard deviation. Data refer to
  simulations at resolution $N=1024$. In the inset, mean square
  vorticity $\langle w^2 \rangle$ versus the fractal dimension
  deficit, $3-D$.}
\label{fig:4}
\end{figure}
This can be done by considering snapshots of $D=3$ turbulence, and
applying the fractal decimation as an {\it a posteriori} filter. It is
immediate to realize that the effect of the static decimation on the
spectrum is $E^{D}_{st}(k) \sim k^{D-3}\,E^{K41}(k)$, implying that
the geometrical action of the decimation goes in the opposite
direction of the dynamical one.\\ We now consider the dynamical effect
of the fractal Fourier decimation on the small-scale structures, by
focusing on the statistics of the vorticity field in the real
space. In Fig.~\ref{fig:4} we plot the probability density function
(PDF) of the vorticity field, normalized with its standard
deviation. We note that already at $D=2.99$, vorticity fluctuations
have changed their intensity of 1 order of magnitude, despite the
fact that the mean enstrophy is practically unchanged. Even more
remarkable, intermittent fluctuations disappear already at $D=2.8$,
where a quasi-Gaussian vorticity PDF is measured. The transition
towards Gaussianity is better quantified considering the vorticity
kurtosis. In Fig.~\ref{fig:5}, we compare results of the fractally
decimated NS equations, with those obtained from the application of
the {\it a posteriori} static mask on three-dimensional
turbulence. The dynamical decimation makes a very fast transition
towards a Gaussian behaviour, such that at $D=2.98$ the kurtosis has
decreased by 30$\%$, to already approach the Gaussian value at
$D=2.8$. In the case of the{\it a posteriori} static decimation,
vorticity kurtosis assumes the Gaussian value only at $D=2.5$, while
staying almost unchanged in the range $D \ge 2.98$. Such a strong
difference clearly indicates that constraining the dynamics to a
subset of modes is critical for the complete development of
intermittency in real space.\\ In Fig.~\ref{fig:6}, we plot the
kurtosis of the longitudinal velocity increment, $K(\delta v_r)=
\langle (\delta_r {\bf v}\cdot \hat{\bf r})^4\rangle/\langle (\delta_r
        {\bf v}\cdot \hat{\bf r})^2\rangle^2$. Notice the sharp
        transition leading to an almost scale-independent, nonintermittent behavior for $D<3$. Moreover, at fixed fractal
        dimension, the effects of increasing the Reynolds number is to
        further reduce the intermittent corrections.\\
\noindent {\it Conclusions}. Fractal mode-reduction is a new route to
design numerical simulations to tackle the problem of intermittency
and to potentially develop multiscale models of turbulence. The first
non trivial result is the robustness of the energy flux under
modereduction. An inertial range of scales with a constant-flux
solution is observed when $D$ is changed and few Fourier modes
survive, at least in the parameter range investigated here. This
is in agreement with the observation that Galerkin truncations do not
alter the inviscid conservation of quadratic quantities, preserving
the existence of exact scaling solutions for suitable third-order
correlation functions (see appendix of Ref.~\cite{biferale_jfm}). The
Fourier spectrum gets a power law correction that can be predicted by
a dimensional argument.
Second and more striking, small-scale intermittency is quickly reduced
for $D<3$ and it almost vanishes already at $D=2.98$. As a
consequence, the presence or absence of some of the Fourier modes
strongly modify the fluctuations of all the others, suggesting the
possibility that intermittency is the result of percolating dynamical
properties across the whole Fourier lattice
\cite{bustamante}.\\ Because of the spectrum modification, the
  scaling exponent of the second order longitudinal structure function
  becomes $\zeta_2 + (D-3)$, where $\zeta_2$ is the measured in the
  $D=3$ homogeneous and isotropic case. This observation would suggest
  that, for the dimension deficit $3-D < 1$, one may obtain
  corrections to the scaling exponents proportional to $3-D$, and the
  anomalous exponents might be computed perturbatively in the
  dimension deficit. If this is the case, the critical dimension $D_c$
  is estimated as the value of $D$ where the dimensional scaling is
  recovered, namely $\zeta_2 + D_c-3 = 2/3$. It gives $D_c \sim 2.96$
  not far from the value of $D$ at which intermittency is observed to
  vanish in the DNS. However, there is no reason to assume that
  anomalous exponents can be computed perturbatively in $3-D$. In
  fact, as mentioned above, intermittency might well be the result of
  multiscale interactions in Fourier space, needing all degrees of
  freedom to develop. Hence, in the presence of even a tiny
  decimation, NS singular solutions responsible for the anomalous
  scaling no longer exist. This would also explain why we observe,
  Fig.~\ref{fig:6}, a reduction of intermittency by increasing the
  resolution at fixed $D$.
Additionally, phenomenological cascade models \cite{frisch} would be
unable to explain the results, as well. \\ In the light of our
results, Fourier decimation can also be seen as a way to introduce a
control parameter and modify the scaling properties of the system,
similarly to what happens for NS equations stirred by a random,
power-law forcing~\cite{FNS77,FF78,prlnoi,noiNJP04,pandit}. In the
latter case, perturbative or semianalytic calculations
\cite{mou,muratore} give indications on the reasons why anomalous
corrections should cancel out for specific values of the control
pararameter. Also, in Ref.\cite{prlnoi} it is numerically shown that when
the random injection becomes the dominant scaling contribution in the
inertial range, a transition to a Gaussian statistics is observed for
the velocity increments. In the present case, however, the connections
between the observed transition to a Gaussian behavior, and a
possible renormalised perturbation theory are to be explored.\\

We acknowledge useful discussions with S. Musacchio and P. Perlekar,
who collaborated with us in the first part of the work. DNS were done
at CINECA (Italy), within the EU-PRACE Project Pra04, No.806. This work
is part of the activity of the ERC AdG NewTURB, No.339032. We thank
F. Bonaccorso and G. Amati for technical support. We thank the
COST-Action MP1305 for support.

\end{document}